\documentclass[aps,pra,showkeys,showpacs,reprint,superscriptaddress]{revtex4-1}

\usepackage{graphicx}
\usepackage{dcolumn}
\usepackage{bm}
\usepackage{amsmath}
\usepackage{amssymb}
\usepackage[final]{changes}
\usepackage{todonotes}
\usepackage{color}

\newcommand{\ket}[1]{|{#1}\rangle}
\newcommand{\bra}[1]{\langle{#1}|}
\newcommand{\ud}{|{\uparrow\downarrow}\rangle}
\newcommand{\du}{|{\downarrow\uparrow}\rangle}
\newcommand{\dd}{|{\downarrow\downarrow}\rangle}
\newcommand{\uu}{|{\uparrow\uparrow}\rangle}

\newcommand{\ehoch}[1]{\text{e}^{#1}}

\newcommand{\rmc}{{\mathrm{c}}}
\newcommand{\rme}{{\mathrm{e}}}
\newcommand{\rmd}{{\mathrm{d}}}
\newcommand{\rmi}{{\mathrm{i}}}
\newcommand{\rmR}{{\mathrm{R}}}
\newcommand{\rmq}{{\mathrm{q}}}

\begin{document}


\title{Purification and switching protocols for dissipatively stabilized entangled qubit states}

\author{Sven M. Hein}
\email{shein@itp.tu-berlin.de}
\affiliation{Technische Universit\"at Berlin, Institut f\"ur Theoretische Physik, Nichtlineare Optik und Quantenelektronik, Hardenbergstra\ss e 36, 10623 Berlin, Germany}
\affiliation{Department of Electrical Engineering, Princeton University, Princeton, New Jersey 08544, USA}
\author{Camille Aron}
\email{camille.aron@ens.fr}
\affiliation{Department of Electrical Engineering, Princeton University, Princeton, New Jersey 08544, USA}
\affiliation{Laboratoire de Physique Th\'eorique, \'Ecole Normale Sup\'erieure, CNRS, Paris, France}
\affiliation{Instituut voor Theoretische Fysica, KU Leuven, Belgium}
\author{Hakan E. T\"ureci}
\affiliation{Department of Electrical Engineering, Princeton University, Princeton, New Jersey 08544, USA}

\date{\today}

\begin{abstract}
Pure dephasing processes limit the fidelities achievable in driven-dissipative schemes for stabilization of entangled states of qubits. We propose a scheme which, combined with already existing entangling methods, purifies the desired entangled state by driving out of equilibrium auxiliary dissipative cavity modes coupled to the qubits. We lay out the specifics of our scheme and compute its efficiency in the particular context of two superconducting qubits in a cavity-QED architecture, where the strongly coupled auxiliary modes provided by collective cavity excitations can drive and sustain the qubits in maximally entangled Bell states with fidelities reaching 90\% for experimentally accessible parameters. 

\end{abstract}


\maketitle

\section{Introduction}

Control of quantum information distributed over multiple qubits is a problem both of fundamental and applied interest. In the field of superconducting electrical circuits, progress in the design of qubits based on Josephson junctions as well as attention to microwave hygiene have enabled coherent control of multiple long-lived qubits~\cite{Shankar2013,Raftery2014,Mlynek2014,Barends2014,Chow2014,Corcoles2015,Riste2015,McKay2015,Salathe2015,Hacohen-Gourgy2015,Roch2014}. Current gate-based quantum information processing architectures generally rely on transient unitary operations for initialization and gate operations and are subject to decoherence, due to the {\it equilibrium} fluctuations at the ambient temperature of the circuit. Therefore conventional approaches to mitigation of decoherence rely dominantly on minimizing the coupling to uncontrolled environmental modes, and this appears to be a daunting task as circuits become more complex. This situation has prompted the search for alternate strategies. 

In recent years, a number of approaches have been considered to address the problem of the preservation of quantum information residing on multiple qubits. Some of these rely on the exploitation of the low-decoherence subspaces in non-Markovian environments~\cite{Beige2000,Kwiat2000,Bellomo2008,Bellomo2009,Franco2013,Hein2015}. Others, inspired from the so-called bang-bang control of classical mechanical systems and the Zeno effect, have focused on the development of protocols that aim at averaging out the effect of the environment~\cite{Viola1998,Viola1999,Viola1999a,Facchi2004,Gordon2008,Biercuk2009,DeLange2010,West2010,Khodjasteh2013,DArrigo2014,LoFranco2014,Jing2015}. A third category, generally referred to as measurement-based feedback, relies on continuously monitoring and correcting the quantum state of a system using a classical controller~\cite{Wiseman1993,Wiseman1994,Wang2005,Carvalho2007,Carvalho2008,Hill2008,Gillett2010,Hou2010,Sayrin2011,Vijay2012,Riste2012,Campagne-Ibarcq2013,Riste2013,Riste2015}. The implementations of most of these strategies either come with a heavy overhead in terms of the resources that are required (\textit{e.g.} ultra-fast electronics), or most often their increasing complexity prevents the scaling to extended systems.

A promising strategy that is being currently explored is based on driven-dissipative approaches, also referred to as quantum bath engineering~\cite{Poyatos1996}. The key idea behind this approach is to modify the fluctuations experienced by the qubits by coupling them to a non-equilibrium electromagnetic environment carefully crafted by strategically applying microwave drives. The non-unitary dynamics resulting from the balance between these drives and the original decoherence mechanisms can stabilize the register of qubits to a desired entangled state~\cite{Plenio2002,Raimond2001,Bellomo2008, Kraus2008,Verstraete2009,Bishop2009,Huelga2012,Franco2013,Zippilli2013,Nikoghosyan2012,Leghtas2013,Kastoryano2011,Stannigel2012,Reiter2012,Reiter2013,Reiter2015,Kapit2016,Krastanov2015,Mirrahimi2014}. Recent experimental work in superconducting circuits has demonstrated that this basic idea can be used to great advantage in stabilizing a non-trivial quantum state of a single qubit~\cite{Murch2012}, a target entangled state of two qubits~\cite{Shankar2013, Lin2013, Berkeley,Liu2016}, or an entire quantum manifold of states of a logical qubit based on Schr\"odinger cat states of a quantum harmonic oscillator~\cite{Leghtas2015}. While most work so far has focused on register initialization protocols, theoretical proposals have been put forward for dissipative error correction~\cite{Cohen2014} and universal quantum computation as well. 

In Ref.~\cite{Aron2014}, we proposed a modular driven-dissipative approach to realize and sustain Bell states between two distant identical qubits. In this scheme, the qubits are placed in an engineered electromagnetic environment, namely two coupled optical cavities, the modes of which (i) mediate an effective interaction between the qubits~\cite{Loo2013}, (ii) provide a simple way to drive the system with external ac microwave sources, and (iii) constitute a well controlled dissipative environment. In practice, the two-qubit system is brought to the desired Bell state by a cavity-stimulated Raman process~\cite{Sweeney2014,Baden2014}. An extended version of this protocol has been shown to scale well with the number of qubits for the stabilization of certain classes of entangled states \cite{Aron2016}. The resource-efficiency of such a simple entangling protocol has recently received experimental evidence~\cite{Berkeley} and arbitrarily long-lived Bell states were obtained with fidelities in excess of 70\%. A closer look at the factors limiting the achievable fidelity in these experiments reveals that pure dephasing processes have a significant role to play. The goal of this work is to propose resource-efficient purification protocols, applicable to any fabricated structure, to reduce the impact of dephasing processes on achievable fidelities. While for concreteness we present our scheme in the context of the superconducting two-qubit system studied in Refs.~\cite{Aron2014, Berkeley}, our considerations apply as well to other cavity QED architectures that permit the placement and coherent control of multiple artificial atoms~\cite{Reithmaier2004,Reitzenstein2006,Vasconcellos2011,Albert2013,Reitzenstein2010,Wickenbrock2013}. Compared to other protocols to reduce dephasing, the main advantage of the presented driven-dissipative approach is its very low cost in terms of required resources and added complexity, making it amenable to extended qubit registers. For instance, combined with the existing entangling method of Ref.~\cite{Aron2014}, it works by simply adding a single additional frequency in the cavity drives.

We first briefly outline the underlying idea behind the stabilization schemes discussed in this work. The Hilbert space of two identical qubits is spanned by the triplet states $\ket{T_-}=\ket{\downarrow\downarrow}$, $\ket{T_0}=\big(\ud+\du\big)/\sqrt{2}$, $\ket{T_+}=\ket{\uparrow\uparrow}$, and the singlet state $\ket{S}=\big(\ud-\du\big)/\sqrt{2}$. We assume that the qubits are initially prepared in a mixture of the one-excitation states $\ket{T_0}$ and $\ket{S}$. For non-interacting qubits, these two states form a degenerate manifold.  Maintaining the qubit system in that subspace can be achieved by several methods, but this is not the main focus of this paper. Here, the goal is to purify the state of the system to bring it, say, to the $\ket{T_0}$ state with a negligible overlap with $\ket{S}$. The scheme consists in (i) lifting the degeneracy between $\ket{T_0}$ and $\ket{S}$, and (ii) applying a drive to depopulate the singlet state in favor of the $\ket{T_0}$ state. Both steps can be performed simultaneously by coupling the qubits to common dissipative degrees of freedom that (i) mediate an effective qubit-qubit interaction hence lifting the degeneracy by $\delta = E_{S} - E_{T_0} \neq 0$, (ii) can be conveniently excited by external drives, and (iii) provide a sharply peaked photonic density of states which will be utilized to single out the favored transition $\ket{S}\rightarrow\ket{T_0}$ in a resonant Raman scattering process, while keeping all the other transition rates orders of magnitudes lower.
For this common environment, we have in mind the photonic modes of optical cavities driven by lasers or microwave sources but, depending on the architecture, they can be any well controlled modes such as plasmonic, mechanical, etc. Let us point out that due to $\delta\neq 0$, the protocol aims at purifying the state of the system to $\ket{T_0}$ or $\ket{S}$, but not at purifying it to an arbitrary superposition of these states.

In the following sections, we lay out this anti-dephasing scheme in the context of a cavity-QED architecture, by combining it with the two-qubit entanglement generation method presented in Ref.~\cite{Aron2014}. After introducing the model for the cavity-qubit system and its dynamics in Sect.~II, Sect.~III is devoted to deriving a reduced \textit{time-dependent} theory for the qubits only and discussing the different dephasing channels. After a series of controlled approximations, the relevant transition rates that govern the non-equilibrium dynamics of the qubits are estimated by means of time-dependent perturbation theory, allowing a detailed presentation of the mechanisms underlying our anti-dephasing scheme.
In Sect.~IV, we go one step further in the analytic description of the scheme by deriving a \textit{time-independent} formulation of the steady-state dynamics, bypassing the transient dynamics. The results corroborate those obtained with time-dependent perturbation theory and provide a much simpler access to numerical solutions for the total cavity-qubit system. In Sect.~V, we present the numerical results that demonstrate the efficiency of the anti-dephasing scheme. We perform two different types of numerical simulations, obtained after two different levels of approximations, which both confirm the validity of the analytic approaches discussed in Sect.~III and IV. Using realistic parameters, we predict improved fidelities for the generation of Bell states with respect to the single frequency scheme of Ref.~\cite{Aron2014}. We conclude our work in Sect.~VI by discussing how this scheme can be scaled up to many-qubit systems for the purification of entangled states on extended networks.

\section{System of two qubits in coupled driven-dissipative optical cavities}
\begin{figure}
\includegraphics[width=\linewidth]{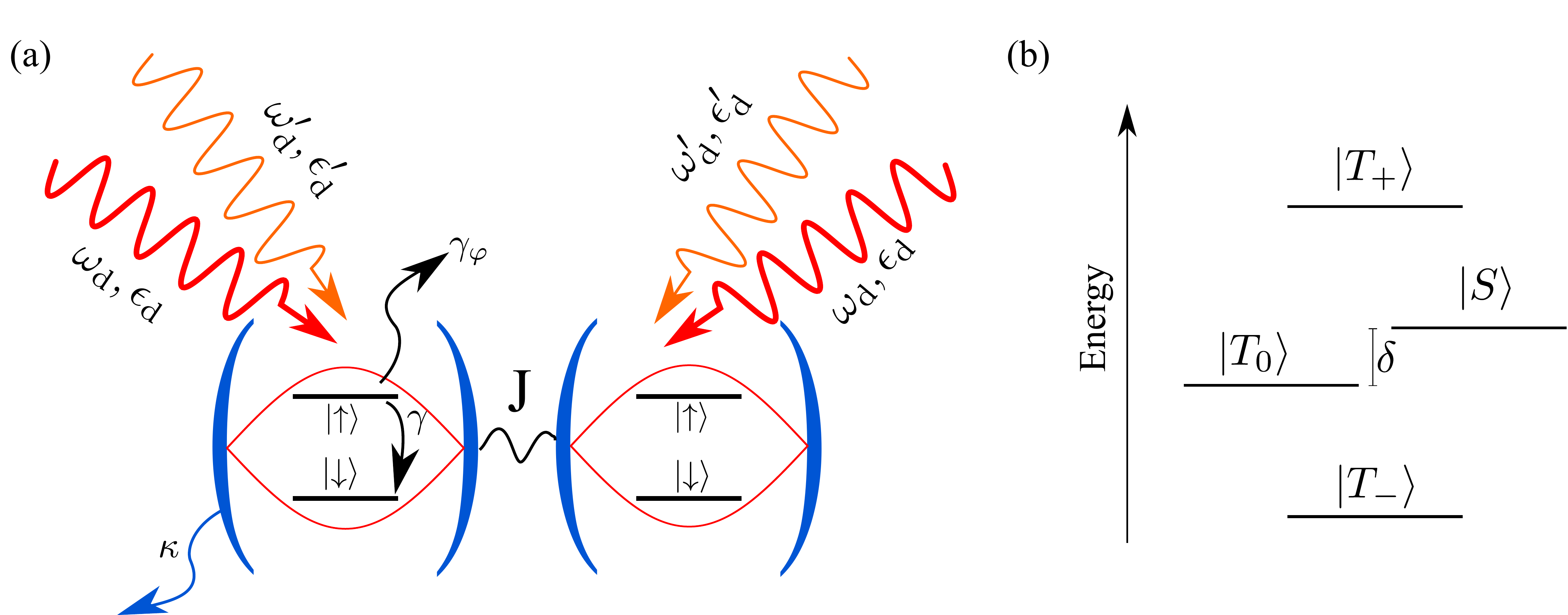}
\caption{\label{fig:system}(a) Implementation of the anti-dephasing scheme in a circuit-QED architecture: two identical qubits are placed in two distinct but coupled single-mode optical cavities. The symmetric optical mode is pumped by two ac drives with different amplitudes and frequencies, one at ($\omega_\rmd, \epsilon_\rmd$), and the other at ($\omega_\rmd', \epsilon_\rmd'$).
$\kappa$ and $\gamma$ are respectively the cavity and qubit decay rates. $\gamma_\varphi$ is the pure dephasing rate, which depends strongly on the photon-mediated energy splitting $\delta = 2 J (g/\Delta)^2$.
(b) Energy spectrum and eigenstates of the coupled qubits.}
 \end{figure}
 
Our system consists of two two-level systems with energy splitting $\omega_\rmq$ (we set $\hbar=1$). They are housed in two coupled single-mode optical cavities with frequency $\omega_\rmc$. Without loss of generality we take the detuning $\Delta \equiv \omega_\rmq - \omega_\rmc > 0$. The two cavities are coupled together, allowing the exchange of photons between the two. Each cavity is driven by two classical ac drives at frequencies $\omega_\rmd$ and $\omega_\rmd'$, and amplitudes $\epsilon_\rmd$ and  $\epsilon_\rmd'$. See also Fig.~\ref{fig:system}(a). The total Hamiltonian reads
\begin{align}
H(t)=& \sum_{i=1}^2 \omega_\rmq \frac{\sigma_i^z}{2}+ \sum_{i=1}^2 g \, \sigma_i^x (a_i^{\phantom{\dagger}}+a_i^\dagger)  \nonumber \\
&
+\sum_{i=1}^2\omega_\rmc a_i^\dagger a_i^{\phantom{\dagger}} -J (a_1^\dagger a_2^{\phantom{\dagger}}+a_2^\dagger a_1^{\phantom{\dagger}})
\nonumber \\ 
&+2\epsilon_\rmd \cos({\omega_\rmd t})\sum_{i=1}^2 (a_i^{\phantom{\dagger}}+a_i^\dagger)   \nonumber \\
&+2\epsilon_\rmd' \cos({\omega'_\rmd t})\sum_{i=1}^2 (a_i^{\phantom{\dagger}}+a_i^\dagger) . \label{eq:H}
\end{align}
The two-level systems are described by the usual Pauli operators $\sigma_{i}^{x,y,z}$ with the cavity index $i=1,2$. The corresponding raising and lowering operators are given by $\sigma_i^\pm \equiv (\sigma^x_i \pm \rmi \sigma^y_i)/2$. $g > 0$ is the light-matter coupling.
We operate in the large detuning regime $\Delta \gg g$ (in practice $g/\Delta \sim 10^{-1}$) and at sufficiently weak drive
amplitudes to ensure the presence of very few photons in the cavities, so that the Schrieffer-Wolff perturbation theory to be applied shortly is well-justified.  
The cavity modes are described by bosonic creation and annihilation operators, $a_i^\dagger$ and $a_i$. $J$ is the tunneling amplitude between the two cavities resulting in two normal modes: a symmetric one $A\equiv(a_1+a_2)/\sqrt{2}$ with frequency $\omega_\rmc^+=\omega_\rmc-J$, and an antisymmetric one $a\equiv(a_1-a_2)/\sqrt{2}$ with frequency $\omega_\rmc^-=\omega_\rmc+J$~\footnote{Note that the notation $\omega_\rmc^\pm=\omega_\rmc \mp J$ differs from that of Ref.~\cite{Aron2014} which reads $\omega_\rmc^\pm=\omega_\rmc \pm J$.}. The symmetric application of each ac drive couples to the symmetric mode, hence effectively pumping symmetric $A$ photons. The first drive with ($\omega_\rmd, \epsilon_\rmd$) is used to pump the collective qubit system into a superposition of singlet $\ket{S}$ and triplet $\ket{T_0}$ Bell states. The second drive with ($\omega_\rmd', \epsilon_\rmd'$) is used to purify the mixture to the desired Bell state and to protect it from dephasing. Elucidating its role is the main purpose of this paper. 

\paragraph*{Dissipative mechanisms ---}
The system is not perfectly isolated and both qubits and cavities are subject to dissipative mechanisms created by their respective environments. Photon loss from each cavity mode occurs at a rate $\kappa$. The excited state $|{\uparrow}\rangle_i$ of each qubit spontaneously relaxes to the ground state $|{\downarrow}\rangle_i$ at a rate $\gamma$. Furthermore, the qubits experience pure dephasing at a rate $\gamma_\varphi$, which will be discussed in detail in Sect.~\ref{par:dephasing}.
In Table I, we gather the typical energy scales. They are closely inspired from the experimental parameters of Ref.~\cite{Berkeley} for superconducting transmon qubits in a three-dimensional (3D) microwave cavity architecture. They obey the hierarchy
$\Delta \gg g, J \gg \kappa \gg \gamma \gg \gamma_\phi$ that justifies the different layers of approximations we shall perform below to construct our analytic approach.

\begin{table}
\begin{ruledtabular}
\begin{tabular}{ccccccc}
$\omega_\rmc=6$&$\omega_\rmq=7$&$g=10^{-1}$&$J=10^{-1}$&$\kappa=10^{-4}$&$\gamma=10^{-5}$                                                                                                                                                                                                                                                                                                                                                                                                                      \end{tabular} 
\end{ruledtabular}\caption{Values (in units of $2\pi$~GHz) used for numerical simulations, unless specified otherwise.}\label{tab:values}  
\end{table}

\paragraph*{Dynamics ---}
The evolution of the system driven by the time-dependent drives and subject to these non-unitary dissipative processes is well described by the following quantum master equation on the system density matrix $\rho(t)$:
\begin{align}
\partial_t \rho(t)  = \mathcal{L}(t) \rho(t)\,, \label{eq:master}
\end{align}
with the time-dependent Liouvillian super-operator
\begin{align}
 \mathcal{L}(t)\rho  = &- \rmi [H(t); \rho] + \kappa \, \mathcal{D}[A]\rho  + \kappa \, \mathcal{D}[a]\rho  \nonumber
\\ 
&  + \sum_i  \gamma \, \mathcal{D}[\sigma^-_i]\rho + \frac{\gamma_\varphi}{2} \, \mathcal{D}[\sigma_i^z]\rho\,,
\end{align}
and where we introduced the Lindblad-type dissipators $\mathcal{D}[X]\rho \equiv \left(X\rho X^\dagger -X^\dagger X \rho + \mbox{H.c.}\right)/2$.

\section{Reduced effective theory for the qubits }
We first present our anti-dephasing scheme at the level of an effective non-equilibrium theory for the qubit subsystem, \textit{i.e.} after integrating out the photonic degrees of freedom. This reduced description of the problem will yield simple dynamical equations governing the populations of the qubit eigenstates. The explicit expressions of the corresponding transition rates will elucidate the  mechanism underlying the scheme.

\paragraph*{Treatment of the light matter coupling ---}
In order to eliminate the light-matter interaction, we use a second-order perturbation theory in $g/\Delta$ by applying a Schrieffer-Wolff transformation ${H} \mapsto \widetilde{H} \equiv \rme^X H \rme^{X^\dagger}$ with
\begin{align}
X\equiv\frac{g}{\sqrt{2}}\left[\frac{A(\sigma_1^++\sigma_2^+)}{\omega_\rmq-\omega_\rmc^+}+\frac{a(\sigma_1^+-\sigma_2^+)}{\omega_\rmq-\omega_\rmc^-}-\text{H.c.}\right]\,.
\end{align}
To minimize the explicit time dependence of the Hamiltonian, we move to a frame rotating at the first drive frequency, $\omega_\rmd$. We then perform a rotating-wave approximation by discarding all terms rotating at $2\omega_\rmd$ and $\omega_\rmd+\omega_\rmd'$. However, we keep terms rotating at $\omega_\rmd-\omega_\rmd'$. This approximation is valid as long as $\omega_\rmd'$ is close to $\omega_\rmd$. We will later find that $\omega_\rmd'$ and $\omega_\rmd$ will have to differ by about $\Delta/2$ for optimal conditions, which is about one order of magnitude lower than $\omega_\rmd$ itself. 

\paragraph*{Linearized photon spectrum ---} In order to eliminate the remaining non-linearities of the type $\Delta (g/\Delta)^2 A^\dagger A \sigma^z_i$ in the Hamiltonian $\widetilde{H}$, we decompose the photon fields into classical mean fields plus quantum fluctuations:
\begin{align}
A\equiv \overline{A}(t) +D \mbox{  and  }  a \equiv \overline{a} + d\,, \label{eq:defineD}
\end{align}
with
\begin{align}
\overline{A}(t)  = \overline{A}_\rmd + \rme^{-\rmi(\omega_\rmd'-\omega_\rmd)t} \overline{A}_\rmd' \mbox{  and  }
\overline{a}  = 0\,, \label{eq:defineoverline}
\end{align}
where the zero above is a consequence of the drives not coupling to the anti-symmetric mode. To lowest order in $g/\Delta$,
\begin{align}
\overline{A}_\rmd \simeq \frac{\sqrt{2}\,\epsilon_\rmd}{\omega_\rmd-\omega_\rmc^++\rmi\kappa/2} \mbox{ and }
\overline{A}_\rmd' \simeq \frac{\sqrt{2}\,\epsilon_\rmd'}{\omega_\rmd'-\omega_\rmc^++\rmi\kappa/2}. \label{eq:DefineBar}
\end{align}
We also define the mean-field photon numbers induced by each drive, $\overline{N}_\rmd\equiv|\overline{A}_\rmd|^2$ and $\overline{N}_\rmd' \equiv|\overline{A}_\rmd'|^2$.

After inserting the identities (\ref{eq:defineD}) into the Hamiltonian $\widetilde{H}$, we neglect those interacting terms that are quadratic in the photon fluctuations and couple to the qubits, \textit{i.e.} of the type $\Delta (g/\Delta)^2 D^\dagger D \sigma_i^z$. In practice, those small terms can be interpreted as photon-fluctuation dependent renormalizations of the qubit frequencies and neglecting them slightly shifts the energetics of the qubits but it has no impact on the mechanism we discuss~\cite{Aron2014}. The resulting Hamiltonian $\widetilde{H}$ can be split up into a part describing the qubits, $\widetilde{H}_{\text{S}}$, a bath part $\widetilde{H}_{\text{B}}$ describing cavity photon fluctuations, and $\widetilde{H}_{\text{S-B}}$ describing the interaction of the qubits with those fluctuations
\begin{align}
 \widetilde{H}_{\text{S}}=&\sum_{i=1}^2 \boldsymbol{h}(t) \cdot \frac{\boldsymbol{\sigma}_i}{2}-\frac{1}{2} J \! \left(\frac{g}{\Delta}\right)^2 \!\! \left(\sigma_1^x\sigma_2^x+\sigma_1^y\sigma_2^y\right) \label{eq:hsys}\\ 
 \widetilde{H}_{\text{B}}=&\left(\omega_\rmc^+-\omega_\rmd\right)D^\dagger D +\left(\omega_\rmc^--\omega_\rmd\right)d^\dagger d \label{eq:hres}\\
 \widetilde{H}_{\text{S-B}}=&\frac{1}{2}\left(\frac{g}{\Delta}\right)^2 \left[\Delta\overline{A}(t)^*+\frac{\epsilon_\rmd}{\sqrt{2}}+\frac{\epsilon_\rmd'}{\sqrt{2}}\rme^{\rmi(\omega_\rmd'-\omega_\rmd)t}\right]\nonumber \\
 &\times \left[D\left(\sigma_1^z+\sigma_2^z\right)+d\left(\sigma_1^z-\sigma_2^z\right)\right]+\text{H.c.}\,.\label{eq:hsysres}
 \end{align}
Above, we introduced an effective time-dependent pseudo-magnetic field $\boldsymbol{h}(t)\equiv \big(h^x(t),h^y(t),h^z(t) \big)$ which is mostly oriented along the $z$ direction and 
\begin{align}
h^x(t)={}&\frac{2g}{\Delta}\Big(\epsilon_\rmd+\epsilon_\rmd'\cos[(\omega_\rmd'-\omega_\rmd)t]\Big)\label{eq:hx}\\
h^y(t)={}&\frac{2g}{\Delta}\Big(\epsilon_\rmd'\sin[(\omega_\rmd'-\omega_\rmd)t]\Big)\\
h^z(t)={}&\omega_\rmq-\omega_\rmd +\left(\frac{g}{\Delta}\right)^2\Big\{ \Delta \nonumber\\
 & \!\! + \Delta(\overline{N}_\rmd+\overline{N}_\rmd'+2\text{Re}[\overline{A}_\rmd^*\overline{A}_\rmd'\rme^{-\rmi(\omega_\rmd'-\omega_\rmd)t}]\big) \label{eq:hz}  \\
 & \!\!  +\sqrt{2}\text{Re}[ \epsilon_\rmd\overline{A}(t)+ \epsilon_\rmd' \rme^{\rmi(\omega_\rmd'-\omega_\rmd)t} \overline{A}(t)]\Big\}. \nonumber
\end{align}

\paragraph*{Qubit spectrum ---}
The term in $J(g/\Delta)^2$ present in $\widetilde{H}_{\text{S}}$ reveals the  effective photon-mediated coupling of the qubits. It lifts the degeneracy of the one-excitation manifold into the maximally entangled states $\ket{T_0}=(\ud+\du)/\sqrt{2}$ and $\ket{S}=(\ud-\du)/\sqrt{2}$, see Fig.~\ref{fig:system}(b). This lifting of degeneracy is crucial since it allows us to differentiate  the two states, which is one of the main ingredients in our anti-dephasing scheme. 
The photonic environment also shifts the qubit eigenenergies. In the absence of any external drive, $\epsilon_\rmd=\epsilon_\rmd'=0$, the eigenstates of $\widetilde{H}_{\text{S}}$ and their respective eigenenergies are, up to second order in $g/\Delta$,
\begin{align}
 \ket{T_+}&=\uu & E_{T_+}&= \Delta_\rmq +\frac{g^2}{\Delta} \nonumber\\
 \ket{S}&= (\ud-\du)/\sqrt{2} & E_S&=\delta/2 \nonumber\\
 \ket{T_0}&= (\ud+\du)/\sqrt{2} &E_{T_0}&=-\delta/2\nonumber\\
 \ket{T_-}&=\dd &E_{T_-}&=- \Delta_\rmq - \frac{g^2}{\Delta} . \label{eq:eigen}
\end{align}
We introduced $\Delta_\rmq \equiv \omega_\rmq-\omega_\rmd$ and the energy splitting  $\delta\equiv2J(g/\Delta)^2$. In the presence of finite driving, the eigenenergies of $\ket{T_\pm}$ acquire small corrections on the order of $(g/\Delta)^2 \epsilon_\rmd^2 / \Delta$, see Eq.~(\ref{eq:hz}).

\subsection{First drive: populating the one-excitation manifold}
Let us for the moment neglect all the terms that stem from the second ac-drive, $\textit{i.e.}$ set $\epsilon_\rmd'=A_\rmd'=0$. From the point of view of the qubit subsystem, the photon fluctuations $D$ and $d$ can be seen as two independent non-interacting baths that are weakly coupled to the qubits, see \textit{e.g.} the prefactor $(g/\Delta)^2$ of the $D^\dagger \sigma_i^z$ terms in Eq.~(\ref{eq:hsysres}). The density of states of these baths is (in the laboratory frame)
\begin{align}
\rho_\pm(\omega)=-\frac{1}{\pi}\text{Im}\frac{1}{\omega-\omega_\rmc^\pm+\rmi\kappa/2}.\label{eq:DOS}
\end{align}
The baths trigger non-equilibrium transitions of the qubits between two possible eigenstates $\ket{k}$ and $\ket{l}$ of $\widetilde{H}_{\text{S}}$ at a rate $\Gamma_{k \to l}$. Assuming that these baths equilibrate with the surrounding zero-temperature environment and following the analysis of Ref.~\cite{Aron2014}, one can use a quantum master equation approach to integrate those weakly-coupled non-interacting degrees of freedom and  derive the corresponding transition rates for the qubits, directly in the steady state, bypassing the transient dynamics. For symmetry reasons, transitions from $\ket{T_-}$ to $\ket{T_0}$, as well as from $\ket{T_0}$ to $\ket{T_+}$ create an excitation in the symmetric mode. 
In contrast, transitions from $\ket{T_-}$ to $\ket{S}$, as well as from $\ket{S}$ to $\ket{T_+}$ create an excitation in the anti-symmetric mode.
A detailed derivation of the transition rates is presented in the appendix. For the transition $\ket{T_-}\rightarrow\ket{T_0}$ as well as for $\ket{T_0}\rightarrow\ket{T_+}$, the rates are 
\begin{equation}
\Gamma_{k\rightarrow l}= 8 \pi \left|\frac{g^3 \epsilon_\rmd^2 }{\Delta^3 \Delta_\rmq}\left(\frac12 + \frac{\Delta}{\omega_\rmd-\omega_\rmc^+}\right) \right| ^2 \rho_+(E_k-E_l+\omega_\rmd).\label{eq:gammatrans}
\end{equation}
where the eigenenergies $E_k$ and $\Delta_\rmq$ are slightly renormalized by the drive, see the Appendix.
The transition rates from $\ket{T_-}\rightarrow\ket{S}$ as well as for $\ket{S}\rightarrow\ket{T_+}$ can be calculated similarly by using $\rho_-$ instead of $\rho_+$.

Let us briefly discuss the driving scheme to $\ket{T_0}$. An analogous argument can be made for targeting $\ket{S}$ instead of $\ket{T_0}$.
In order to populate efficiently the state $\ket{T_0}$   starting from the ground state $\ket{T_-}$, one maximizes the rate $\Gamma_{T_-\rightarrow T_0}$ by choosing $\omega_\rmd$ such that the photonic density of states $\rho_{+}$ in Eq.~(\ref{eq:gammatrans}) is probed at its maximum $2/\pi\kappa$. The optimal value is achieved for $\omega_\rmd=\omega_\rmc^++E_{T_0}-E_{T_-}$, which is approximately equivalent to $2\omega_\rmd\simeq\omega_\rmq+\omega_\rmc^+$. For high finesse cavities this is a highly selective process with rate~\cite{Aron2014}
\begin{align}
\Gamma_{T_-\rightarrow T_0} \simeq 400 \frac{g^6 \epsilon_\rmd^4}{\Delta^8 \kappa}.\label{eq:ratedrive}
\end{align}

\paragraph*{Dephasing mechanisms ---} \label{par:dephasing}
No matter how strongly this transition is excited, there are two finite dephasing channels which mix the Bell states $\ket{T_0}$ and $\ket{S}$ into one another, severely limiting the maximally achievable purity of the target steady state. First, as we already mentioned earlier, the environment is responsible for mediating such transitions at the so-called pure dephasing rate $\gamma_\varphi$. Note that once the two qubits are effectively coupled via the cavity-mediated interaction and the degeneracy between singlet and triplet state is lifted by $\delta$, $\gamma_\varphi$ can be greatly suppressed because this bath-mediated process involves probing the bath spectrum at a finite frequency $\delta$ rather than at zero frequency. In the system discussed in Ref.~\cite{Aron2014}, there are strong experimental indications~\cite{Berkeley} that once coupled, the  qubit dephasing rate is indeed reduced by at least a factor of 10 with respect to the single qubit rate. This is reflected in our choice of the numerical range of $\gamma_\varphi$ in our simulations, which we estimate to be between $1.0\times 2 \pi$~kHz and $10.0\times 2 \pi$~kHz based on recent experiments~\cite{Berkeley}. 
Second, there is an additional dephasing channel mixing $\ket{S}$ and $\ket{T_0}$ that arises purely due to non-equilibrium conditions, in which the state $\ket{T_+} \equiv \ket{\uparrow\uparrow}$ participates as an intermediate state: $\ket{T_0}\rightarrow\ket{T_+}\rightarrow\ket{S}$. Indeed, when driving from $\ket{T_-}$ to $\ket{T_0}$, one also drives off-resonantly transitions from $\ket{T_0}$ to $\ket{T_+}$. 
In the steady state, the corresponding effective dephasing rate $\Gamma_{T_0\rightarrow S}^{\text{eff}}$ of this transition can be estimated as~\cite{Aron2014}
\begin{equation}
\Gamma_{T_0\rightarrow S}^{\text{eff}}= \frac{25 g^2 \epsilon_\rmd^4 \kappa}{2\Delta^4 J^2}\,.
\end{equation}
For the values specified in Tab.~\ref{tab:values}, and for the typical value of $\epsilon_\rmd=0.1 \times 2 \pi$~GHz, we find $\Gamma_{T_0\rightarrow S}^{\text{eff}}\approx 0.1 \times 2 \pi$~kHz. However, for larger $\kappa$ and $\epsilon_\rmd$, as discussed below in the paragraph on the ``switching scheme'', this rate can easily get into regions of $10 \times 2 \pi$~kHz. These values are comparable to our estimates of $\gamma_\varphi$.
 Below, we present a simple non-equilibrium route, involving a second carefully tuned microwave drive, to counteract both these dephasing channels.

\subsection{Second drive: Anti-dephasing scheme}
We now discuss a route to reduce the impact of both of these dephasing mechanisms by the use of the second ac drive to deplete the population of the undesired state and improve the purity of the target steady state. Let us now analyze those time-dependent terms in the Hamiltonian~(\ref{eq:hsys})-(\ref{eq:hsysres}) which are due to the second drive.
We can safely neglect the terms in $\epsilon_\rmd' \cos((\omega_\rmd'-\omega_\rmd)t)\left(\sigma_1^x+\sigma_2^x\right)$ and $\epsilon_\rmd' \sin((\omega_\rmd'-\omega_\rmd)t)\left(\sigma_1^y+\sigma_2^y\right)$. Indeed, they would drive direct transitions between states with different quantum numbers $m=\langle{\sigma_1^z+\sigma_2^z}\rangle$, however the frequency of our second drive will be tuned far off these resonances by an amount on the order of $\Delta$. We also neglect the term in $D \rme^{\rmi(\omega_\rmd'-\omega_\rmd)t}\left(\sigma_1^z+\sigma_2^z\right) + \mathrm{H.c.}$ since it only couples to $\ket{T_\pm}$ and does not change the state of the qubits. The terms which are relevant for our scheme are
\begin{align}
\frac{1}{2}  \left(\frac{g}{\Delta}\right)^2 \left(\Delta \overline{A}'_\rmd +\frac{\epsilon_\rmd'}{\sqrt{2}}\right) \rme^{\rmi(\omega_\rmd'-\omega_\rmd)t} d \left(\sigma_1^z-\sigma_2^z\right)  + \text{H.c.}.
\end{align}
These system-bath terms drive transitions between $\ket{T_0}$ and $\ket{S}$ and simultaneously change the number of photons in the antisymmetric cavity mode. The corresponding rates can be estimated via the Fermi's Golden rule:
\begin{align}
\Gamma_{T_0\rightarrow S}'&=4 \pi \left| \Lambda \right|^2 \rho_-(E_{T_0}-E_S+\omega_\rmd')\,,  \\
\Gamma_{S\rightarrow T_0}'&=4 \pi \left| \Lambda \right|^2 \rho_-(E_{S}-E_{T_0}+\omega_\rmd')\,, \label{eq:gammapure}
\end{align}
with $\Lambda \equiv \epsilon_\rmd'\left({g}/{\Delta}\right)^2[{1}/{2}+\Delta/(\omega_\rmd'-\omega_\rmc^+)]$ where we simplified the expressions using Eq.~(\ref{eq:DefineBar}). Using the high finesse of the cavities, \textit{i.e.} the sharply peaked nature of $\rho_-$, one can fine-tune the drive frequency $\omega_\rmd'$ to maximize one of the rates above while keeping the other rate orders of magnitude weaker. To enhance, say, the purity of the $\ket{T_0}$ state, the optimal frequency is given by
\begin{align} \label{eq:opt2}
\omega_\rmd'=\omega_\rmc^--\delta\,,
\end{align}
 yielding the rate 
\begin{align}
\Gamma_{S\rightarrow T_0}' \simeq 2\frac{g^4{\epsilon'_\rmd}^2}{\Delta^2 J^2 \kappa}. \label{eq:ratepure}
\end{align}
Analogously, the same rate can be achieved for the enhancement of the purity of $\ket{S}$ when choosing $\omega_\rmd'=\omega_\rmc^-+\delta$.
{This is the main result of the paper:} within the one-excitation subspace of coupled qubits, Bell states can be purified by scattering photons inelastically into a dissipative cavity mode.  
Comparing Eq.~(\ref{eq:ratepure}) with Eq.~(\ref{eq:ratedrive}), we find that only a very weak second drive amplitude, $\epsilon_\rmd' \ll \epsilon_\rmd$, is necessary to achieve the purification process at rates on the same order of magnitude as the transition driven by the first drive. The actions of the two drives are depicted in Fig.~\ref{fig:switching}(a).

Note that the above rates were computed under the assumption that photon fluctuations in the antisymmetric cavity mode are vanishingly small, $\langle{d^\dagger d}\rangle=0$. However this is only an approximation since, even at zero temperature, they can be dynamically populated by the noise photons produced by the two drives through the very processes we discussed above. For a finite population, the reverse transitions which \textit{remove} one noise photon from the cavity mode and \textit{decrease} the purity of the desired state are also present. For small population, the ratio between backward and forward rate can be estimated by
\begin{equation}
 \frac{\Gamma_{\text{backw.}}}{\Gamma_{\text{forw.}}}\simeq\frac{\langle{d^\dagger d}\rangle}{1+\langle{d^\dagger d}\rangle}. \label{eq:fwdbckwd}
\end{equation}
It is therefore desirable to keep the population of antisymmetric photon fluctuations as low as possible. This can be done by driving different cavity modes to trigger transitions to the Bell state subspace (first drive) and the purification process (second drive).  This is always possible since we are not bound to coupling the second drive to the symmetric mode; it can also be coupled it to the antisymmetric mode by driving both cavities with a $\pi$-phase difference. The purification process would then scatter into the symmetric mode.
This also shows that a finite cavity decay $\kappa$ can be beneficial for the process: There is a balance between frequency selectivity (favored by a smaller $\kappa$) and the suppression of the backward process (favored by a larger $\kappa$). 

We emphasize again that the two processes induced by the two drives are independent of each other. In particular, any mechanism to bring the qubits into the subspace spanned by $\ket{T_0}$ and $\ket{S}$ is suitable to be combined with the purification process brought by the drive $(\omega_\rmd',\epsilon_\rmd')$. The only prerequisites for the purification process are (a) a finite energy difference $\delta$ between the Bell states, (b) a photonic mode coupled to the qubits to be pumped by a coherent ac drive, and (c) a second photonic mode with a decay rate $\kappa\ll\delta$.

\subsection{Switching scheme} 
\begin{figure}
\includegraphics[width=\linewidth]{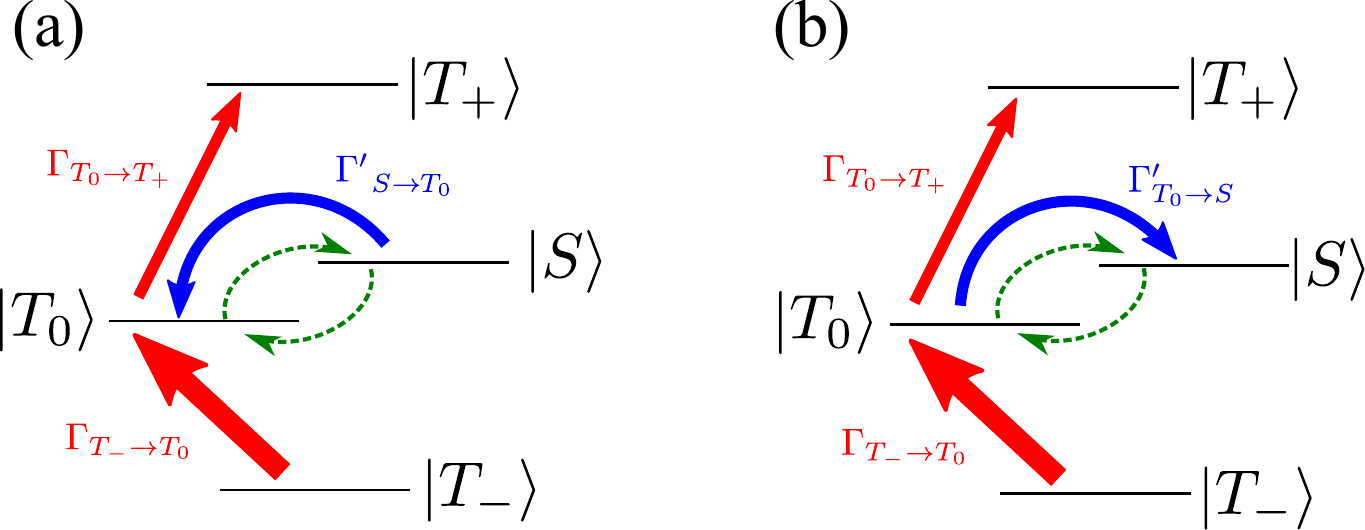}
\caption{\label{fig:switching} (color online) Two different ways to use the second drive.  Thick (red) straight arrows represent the main resonant pumping by the first ac drive, and thinner (red) straight arrows represent represent off-resonant transitions. Round (blue) solid arrows show the action of the second drive. Undesirable dephasing processes are represented with round (green) dashed arrows. (a) Direct scheme: main pump to $\ket{T_0}$, second drive used to reduce dephasing; (b) Switching scheme: main pump to $\ket{T_0}$, second drive used to transfer to $\ket{S}$. 
}
 \end{figure}

The second drive can also be used for switching between the Bell states (\textit{cf.} Fig.~\ref{fig:switching}). For instance, take the first drive as fixed and targeting the $\ket{T_0}$ state as above. 
The second drive may now not only be used to purify this state as in Fig.~\ref{fig:switching}(a), but can also be tuned to completely transfer the population to the $\ket{S}$ state, see Fig.~\ref{fig:switching}(b).
The latter scheme may be beneficial compared to using the first drive to target directly the singlet state because it can considerably reduce the incidental transitions  from $\ket{S}$ to $\ket{T_+}$ by making them more off-resonant from the main transition. While there is still off-resonant driving from $\ket{T_0}$ to $\ket{T_+}$ present when this protocol is applied, these transitions are negligible due to the low population of the $\ket{T_0}$ state. This in turn allows for a stronger pumping and an effective depletion of $\ket{T_-}$. However, stronger pumping also leads to higher intensities in the cavity modes, which also increases the strength of the ``backward process'' discussed above. Our numerical results discussed below in Sect.~V show that the relative performance of the two schemes, direct driving or switching, depend on the precise experimental parameters. For relatively lossy cavities, the switching scheme will be more favorable, since cavity decay results in a larger off-resonant transition rate to $\ket{T_+}$ and an accompanying reduction of the intensity in the cavity modes.

\section{Alternate steady-state master equation approach}
The presence of the two drives introduces two distinct external frequencies in the problem,  $\omega_\rmd$  and  $\omega_\rmd'$. In Sect.~III, we eliminated the explicit time-dependence introduced by the first drive by working in the frame rotating at $\omega_\rmd$ and the effects of the second drive, entering the Hamiltonian \textit{via} time-dependent terms rotating at $\omega_\rmd - \omega_\rmd'$, were tackled by means of time-dependent perturbation theory. We show below that one can formulate an alternate master equation description of the problem in which \textit{both} time-dependencies are fully eliminated from the Liouvillian $\overline{\mathcal{L}}$ and the steady state can be accessed directly, bypassing the transient dynamics.
In practice, the steady-state density matrix $\rho^{\rm NESS}$ will be computed by simply solving $\overline{\mathcal{L}} \rho = 0$. 
We note that such a gauging away of all explicit time dependencies is only possible as long as both processes, driving to the one-excitation subspace and subsequent purification, scatter into \textit{different} modes. In that particular configuration, we can use the fact that for each mode there exists a single relevant frequency, either $\omega_\rmd$ or $\omega_\rmd'$, and neglect all terms rotating with the other frequency.

Let us now derive the corresponding time-independent Hamiltonian in the particular case in which the symmetric mode is driven to bring the qubit system to the one-excitation subspace (\textit{i.e.} targeting $\ket{T_0}$) while the anti-symmetric mode is used to purify the state $\ket{T_0}$. Starting from Eqs.~(\ref{eq:hsys})-(\ref{eq:hsysres}), we perform the additional rotating frame transformation $\widetilde{H}(t) \mapsto U(t)[\widetilde{H}(t)-\rmi\partial_t]U(t)^\dagger$ with $U(t)=\exp[\rmi(\omega_\rmd'-\omega_\rmd)t\, d^\dagger d]$. We then neglect all time-dependent terms that act directly on the spin operators $\sigma_i^x$ and $\sigma_i^y$, since these are not relevant for the transition triggered by the second drive -- the second drive will be off-resonant to any transition changing the number of excitations in the system. We also neglect all the remaining time-dependent terms, since all dominant processes involving the second drive also involve the antisymmetric mode and are therefore now time-independent. This series of approximations leads to the time-independent Hamiltonian
\begin{equation}
 \overline{H}=\overline{H}_{\text{S}}+\overline{H}_{\text{B}} 
 + \overline{H}_{\text{S-B}}, \label{HamiltonianTIndep}
\end{equation}
with
 \begin{align}
\overline{H}_{\text{S}}&=\sum_{i=1}^2 \boldsymbol{h} \cdot \frac{\boldsymbol{\sigma}_i}{2} -\frac{1}{2} J \left(\frac{g}{\Delta}\right)^2 \left(\sigma_1^x\sigma_2^x+\sigma_1^y\sigma_2^y\right) \\
\overline{H}_{\text{B}}&= \left(\omega_\rmc^+-\omega_\rmd\right)D^\dagger D +\left(\omega_\rmc^--{\omega_\rmd'}\right)d^\dagger d \\
\overline{H}_{\text{S-B}}&=\frac{1}{2}\frac{g^2}{\Delta}\left(D^\dagger D+d^\dagger d\right)\left(\sigma_1^z+\sigma_2^z\right)  \label{eq:HamiltonianRot} \\
&+\frac{1}{2}\left(\frac{g}{\Delta}\right)^2\Big[\Big(\overline{A}_\rmd \Delta + \frac{\epsilon_\rmd}{\sqrt{2}}\Big)D\Big(\sigma_1^z+\sigma_2^z\Big)+\text{H.c.}\Big]\nonumber \\
&+\frac{1}{2} \left(\frac{g}{\Delta}\right)^2 \Big[\Big(\overline{A}_\rmd' \Delta + \frac{\epsilon_\rmd'}{\sqrt{2}}\Big)d\Big(\sigma_1^z-\sigma_2^z\Big)+\text{H.c.}\Big],\nonumber
\end{align}
with the effective static magnetic field $\boldsymbol{h}\equiv(h^{x},h^{y},h^{z})$ given by
\begin{align}
h^{x}={}&\frac{2g}{\Delta}\epsilon_\rmd\,,\quad h^{y}={} 0 \,,\\
h^{z}={}&\omega_\rmq-\omega_\rmd +\left(\frac{g}{\Delta}\right)^2\Big[\Delta\big(1+\overline{N}_\rmd+\overline{N}_\rmd'\big)\nonumber \\
 &\qquad \qquad \qquad \qquad  +\sqrt{2}\mathrm{Re}(\epsilon_\rmd \overline{A}_\rmd+\epsilon_\rmd' \overline{A}_\rmd')\Big].
\end{align}
Note that depending whether we aim at an analytic treatment, integrating out the degrees of freedom of the photon fluctuations, or at a numerical integration, we have the choice to neglect or not the terms of the form $D^\dagger D\sigma^z_i$ and $d^\dagger d\sigma^z_i$ in Eq.~(\ref{eq:HamiltonianRot}) above.
In the eigen-basis of $\overline{H}$, the steady-state density matrix is the solution of 
\begin{align}
0= \overline{\mathcal{L}} \rho\,, \label{eq:master2}
\end{align}
with the \textit{time-independent} Liouvillian super-operator
\begin{align}
 \overline{\mathcal{L}}\rho  = &- \rmi [\overline{H}; \rho] + \kappa \, \mathcal{D}[A]\rho + \kappa \, \mathcal{D}[a]\rho \nonumber
\\ 
&  + \sum_i   \gamma \, \mathcal{D}[\sigma^-_i]\rho + \frac{\gamma_\varphi}{2} \, \mathcal{D}[\sigma_i^z]\rho\,.
\end{align}

Similarly to the reduced effective theory for the qubits developed in Sect.~III, this approach relies on Schrieffer-Wolff perturbation theory and a series of rotating wave approximations. As a consistency check, one can verify that Fermi's Golden Rule applied on $\overline{H}$ yields the exact same transition rates as presented in Sect.~III.
However, the great strength of this approach is that the density matrix of the full system (qubits and photons) can now be obtained directly in the steady-state, \textit{i.e.} bypassing the transient dynamics, by simply numerically solving Eq.~(\ref{eq:master2}). This amounts to obtaining steady-state results orders of magnitude faster than a full numerical integration of the original time dynamics given in Eq.~(\ref{eq:master}). A comparison of this approach with the numerical solution of the full (time-dependent) master equation Eq.~(\ref{eq:master}) is presented below in Sect.~V.

\section{Efficiency of the anti-dephasing scheme}

\begin{figure}
 \includegraphics[width=\linewidth]{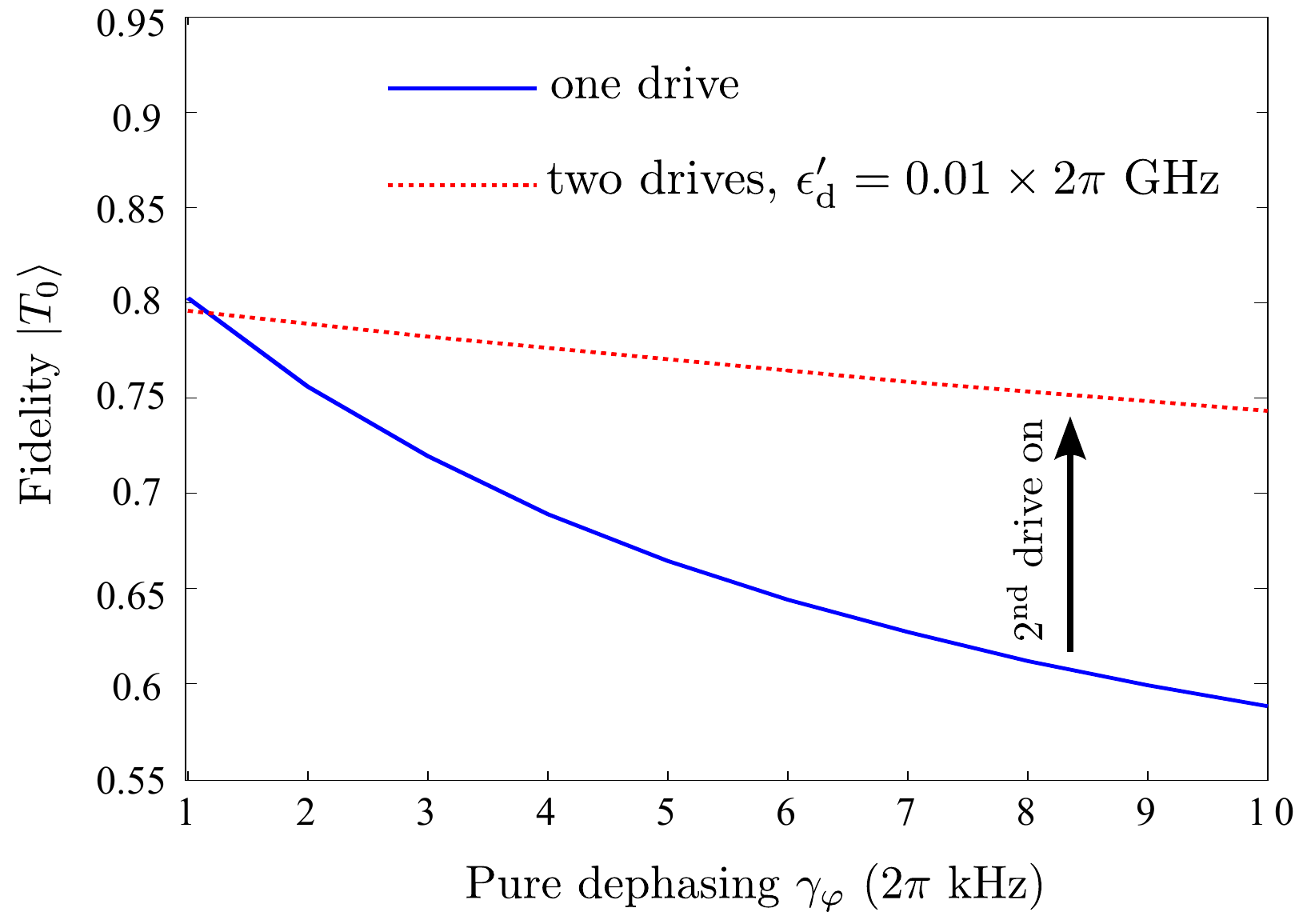}
 \caption{\label{fig:dephasingsweep} (color online) Optimum fidelity to the Bell state $\ket{T_0}$ as a function of the pure dephasing rate $\gamma_\varphi$. We use $\epsilon_\rmd=0.1\times2\pi$~GHz.
Solid, blue line: only one drive is used to drive from $\ket{T_-}$ to $\ket{T_0}$, without employing our proposed purification scheme. The fidelity is greatly suppressed for stronger pure dephasing rates.
Dotted, red line: adding a second drive with strength $\epsilon_\rmd'=0.01\times2\pi$~GHz makes the system much less vulnerable to pure dephasing.
 }
\end{figure}

\subsection{Full time-dynamics results}
We compute the dynamics of the entire system subject to dissipation and the two ac drives by numerically integrating the master equation in Eq.~(\ref{eq:master}) in the frame rotating at $\omega_\rmd$, after having discarded all terms rotating at $2\omega_\rmd$ or $\omega_\rmd+\omega_\rmd'$. We truncate the photon Hilbert spaces to maximum 5 photons in each mode. If not specified otherwise, we use the values specified in Table~\ref{tab:values}. Targeting the qubit state $\ket{X}$, the purity of the steady state is evaluated in terms of the steady-state fidelity, computed as $\mathcal{F}(\ket{X}) = \lim\limits_{t\to\infty} \langle X | \rho_\sigma(t) | X \rangle \in [0,1]$ where $\rho_\sigma$ is the reduced density matrix of the qubit sector.

To demonstrate the efficiency of our scheme to purify the target entangled state, say $\ket{T_0}$, we compute the optimal steady-state fidelities to $\ket{T_0}$ in the presence of the second drive for different realistic values of the pure dephasing rate $\gamma_\varphi$, and compare them to the case in which the second drive is turned off ($\epsilon_\rmd' = 0$). The results are presented in Fig.~\ref{fig:dephasingsweep}.
The qubits are driven to $\ket{T_0}$ with a main drive of amplitude $\epsilon_\rmd=0.1 \times 2\pi$~GHz and optimal frequency $\omega_\rmd \simeq 6.4548\times 2\pi$~GHz. The frequency of the second drive is kept optimally tuned to $\omega_\rmd'=\omega_\rmc^--2J(g/\Delta)^2=6.098\times 2\pi$~GHz, see Eq.~(\ref{eq:opt2}). Note that this frequency, and hence the protocol, is independent of the parameters of the first drive, $\omega_\rmd$ and $\epsilon_\rmd$.
The beneficial effect of the second drive is very clear, especially for cases in which the pure dephasing rate $\gamma_\varphi$  is larger. For example, in the case $\gamma_\varphi = 10 \times 2\pi$~kHz, the Bell state is purified by an additional $15\%$. 
For very small pure dephasing rates $\gamma_\varphi \sim 1 \times 2\pi~$kHz, the apparent negative impact of the second drive in Fig.~\ref{fig:dephasingsweep} can be cured by reducing $\epsilon_\rmd'$ and/or increasing the cavity losses $\kappa$ in order to  minimize the backwards processes described above (cf. Eq.~\ref{eq:fwdbckwd}).

\begin{figure}
 \includegraphics[width=\linewidth]{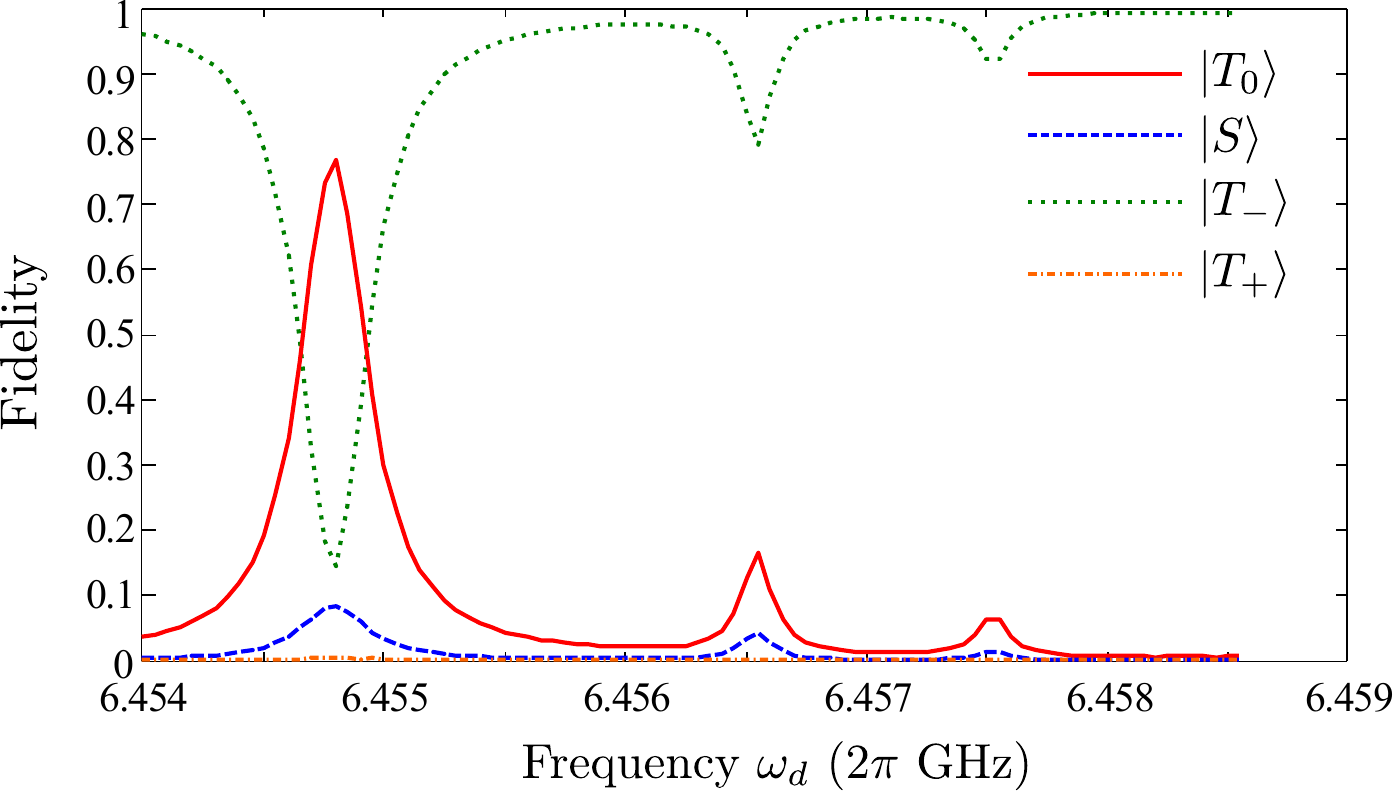}
 \caption{\label{fig:wdsweep} Occupations of $\ket{T_0}$, $\ket{S}$, $\ket{T_-}$, and $\ket{T_+}$ as a function of the frequency $\omega_\rmd$ of the first drive, with the second drive tuned to enhance the fidelity to $\ket{T_0}$.
 The side peaks originate from higher-order photon processes, \textit{cf.} discussion in the text. ($\epsilon_\rmd=0.1, \gamma_\varphi=5.0\times10^{-6}$, and $\epsilon_\rmd'=0.01$ in units of $2\pi$~GHz). 
 }
\end{figure}

In Fig.~\ref{fig:wdsweep}, we study the dependence of the steady-state fidelities on the frequency $\omega_\rmd$ of the first drive once the second drive is on. We set $\gamma_\varphi=5.0\times2\pi$~kHz, $\epsilon_\rmd=0.1\times2 \pi $~GHz, and $\epsilon_\rmd'=0.01\times2 \pi $~GHz. As expected, we find a sizable peak in the fidelity to $\ket{T_0}$ around $\omega_\rmd \simeq (\omega_\rmq+\omega_\rmc)/2$. Moreover, side peaks appear at higher frequencies. 
To understand the mechanism underlying these side peaks, let us first point out that they do not appear in the absence of the second drive ($\epsilon_\rmd'=0$) but show up around $\omega_\rmd \simeq \omega_\rmc^++E_{S}-E_{T_-}$ for very small $\epsilon_\rmd'$. This energy conservation rule transparently indicates that they arise from qubit transitions from $\ket{T_-}$ to $\ket{S}$ with the first drive via the simultaneous emission of a photon into the \textit{symmetric} cavity mode. For larger $\epsilon_\rmd'$, the side peak splits into in several distinct peaks that can only be captured numerically by working with photon number cutoffs larger than one. This indicates that they are the result of higher-order multi-photon processes. 

\subsection{Validation of the steady-state approach}
\begin{figure}
 \includegraphics[width=\linewidth]{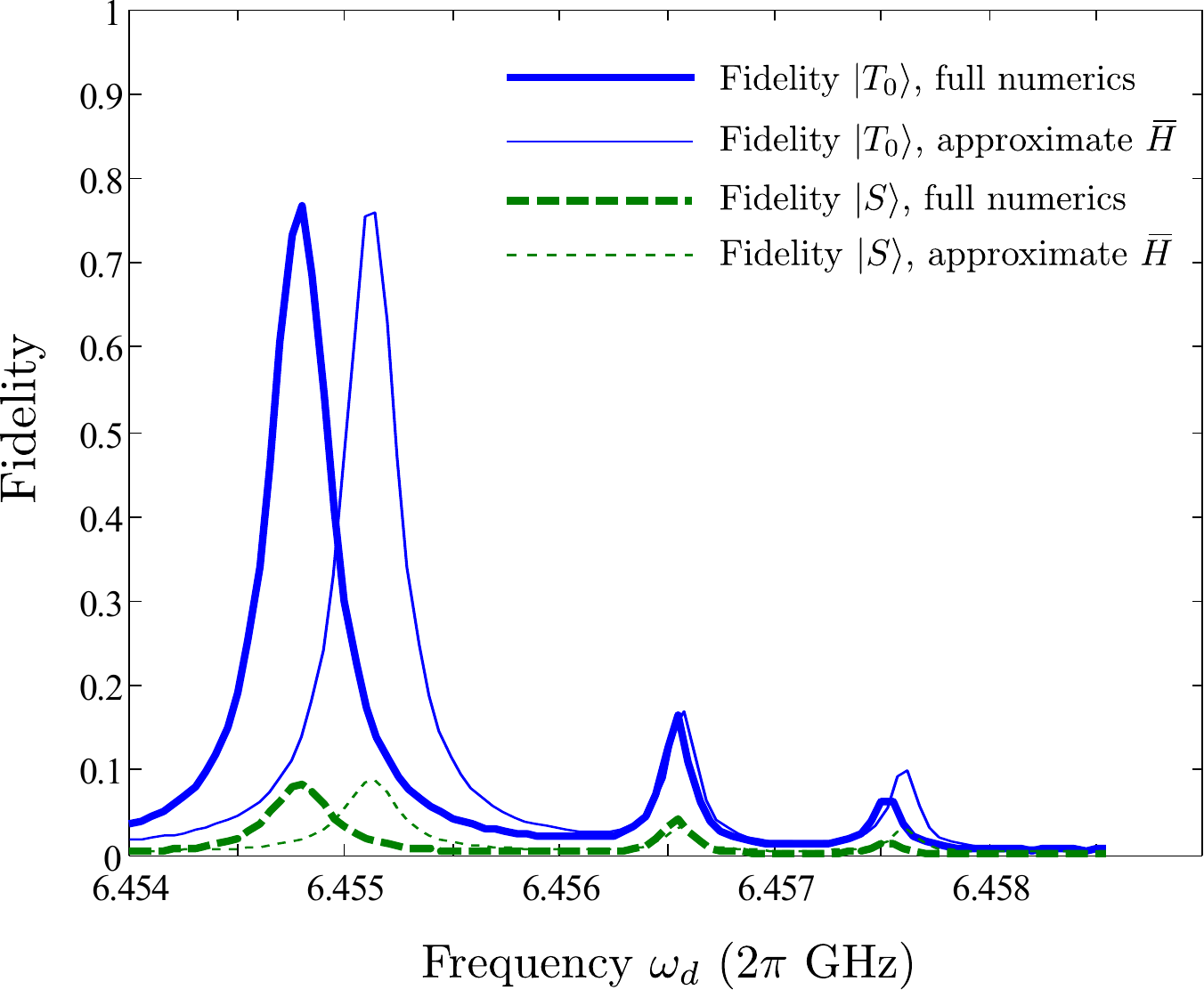}
 \caption{\label{fig:smartrf} Comparison of the steady-state fidelities to generate $\ket{T_0}$ and $\ket{S}$, calculated with (i) the full time-dependent Liouvillian in Eq.~(\ref{eq:master}) and  (ii) the approximate steady-state Liouvillian based on the time-independent Hamiltonian in Eq.~(\ref{eq:HamiltonianRot}). The parameters are identical to those of  Fig.~\ref{fig:wdsweep}. We see that, except for a small energy mismatch, the steady-state formulation reproduces very well the exact results.}
\end{figure}
Having access to the full time-dynamics of the entire system until a steady state is reached, we can check the validity of the steady-state formulation we developed in Sect.~IV. In Fig.~\ref{fig:smartrf}, we compare the results of both methods by plotting the steady-state fidelities to the singlet and triplet states as a function of the main drive frequency. We used the same set of data as we used for Fig.~\ref{fig:wdsweep}. We find that, except from a small energy shift between optimal frequencies, the approximate steady-state formulation reproduces remarkably well the fidelity peaks. These shifts in optimal frequencies derive from the neglected terms that are second order in photon fluctuations such as $D^\dagger D\sigma^z_i$. Thus, the presented quasi-analytic approach is perfectly suitable to make rapid, yet still accurate, predictions of the achievable fidelities.
\subsection{Dependence on the protocol parameters}
\begin{figure}
 \includegraphics[width=\linewidth]{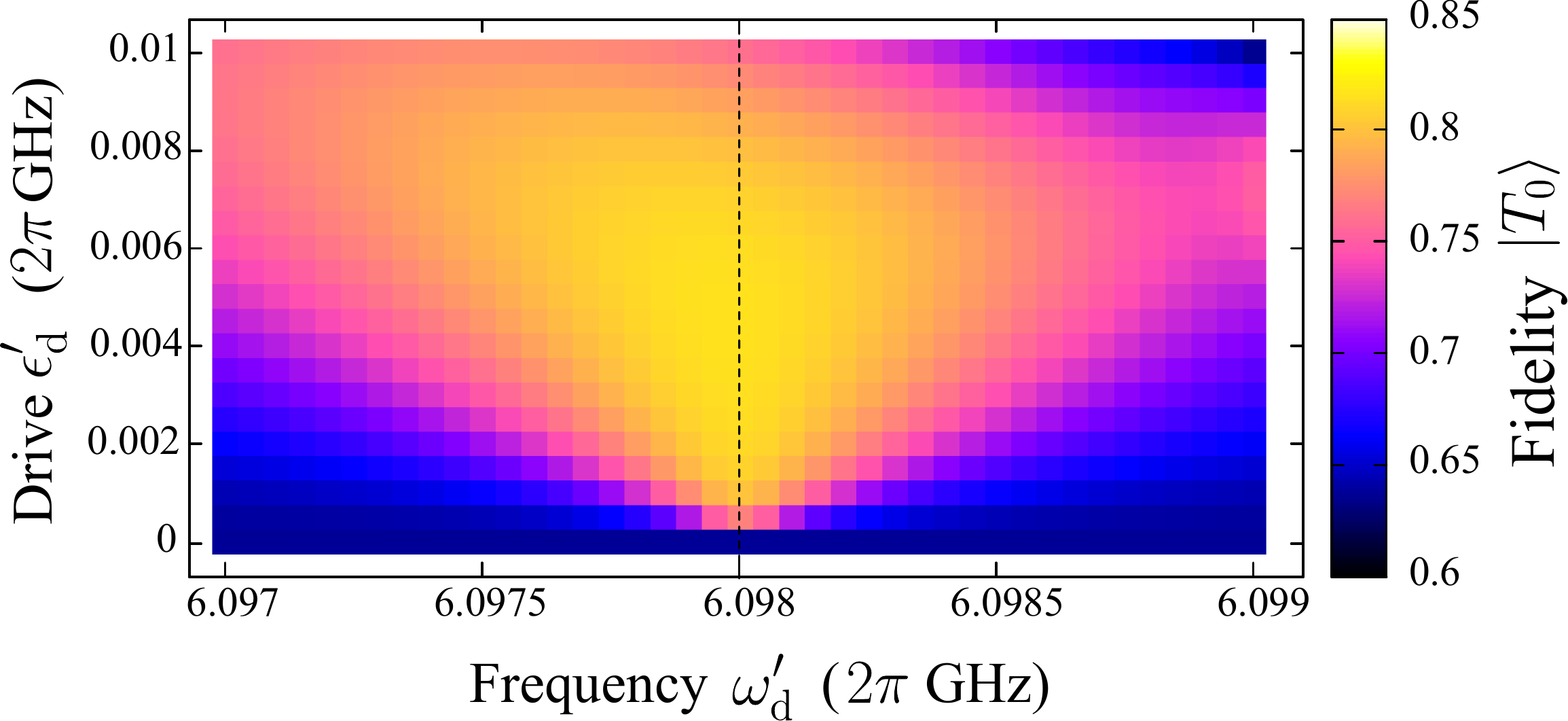}
 \caption{\label{fig:primedsweeps} (color online)
 Fidelity to $\ket{T_0}$ as a function of $\epsilon_\rmd'$ and $\omega_\rmd'$. $\epsilon_\rmd=0.1\times2\pi$~GHz and $\omega_\rmd=6.45515\times 2 \pi$~GHz are set to maximize the fidelity in the absence of the second drive and $\gamma_\varphi=5.0\times 2\pi$~kHz.
The best fidelities are obtained for $\epsilon_\rmd'=0.005 \times 2 \pi$~GHz and $\omega_\rmd' = \omega_\rmc^- - 2(g/\Delta)^2 J=6.098\times 2 \pi$~GHz (dashed line). 
 }
\end{figure}

Having validated the method, we take profit of its numerical simplicity to fully analyze the dependence of the protocol on the second drive parameters. In Fig.~\ref{fig:primedsweeps}, we plot the steady-state fidelity to  $\ket{T_0}$ as a function of $\epsilon_\rmd'$ and $\omega_\rmd'$ while keeping the first drive parameters, $\epsilon_\rmd$ and $\omega_\rmd$, fixed and already optimized to maximize the fidelity in the absence of the second drive. The pure dephasing rate is set to $\gamma_\varphi=5.0\times 2\pi$~kHz which is the mid-value of Fig.~\ref{fig:dephasingsweep}.

We find an increase of the fidelity already for very low second drive amplitude $\epsilon_\rmd' < 10^{-3} \times 2 \pi$~GHz. The optimal value is reached around $\epsilon_\rmd'=5.0\times 10^{-3}\times 2 \pi$~GHz, yielding fidelities $\mathcal{F} \simeq 0.82$ that are even higher than in the above computations which were performed with twice the value of $\epsilon_\rmd'$.
For larger amplitudes, the level of antisymmetric photonic fluctuations is higher and the success of the protocol is hindered by the backward process, as discussed around Eq.~(\ref{eq:fwdbckwd}).
 As expected from Eq.~(\ref{eq:opt2}), the optimal value for $\omega_\rmd'$ is set by $\omega_\rmc^- -2 (g/\Delta)^2J$, and very good results are still obtained as long as $\omega_\rmd'$ is tuned within a distance $\kappa$ of this optimal value.
 
\subsection{Switching scheme}
\begin{figure}
 \includegraphics[width=\linewidth]{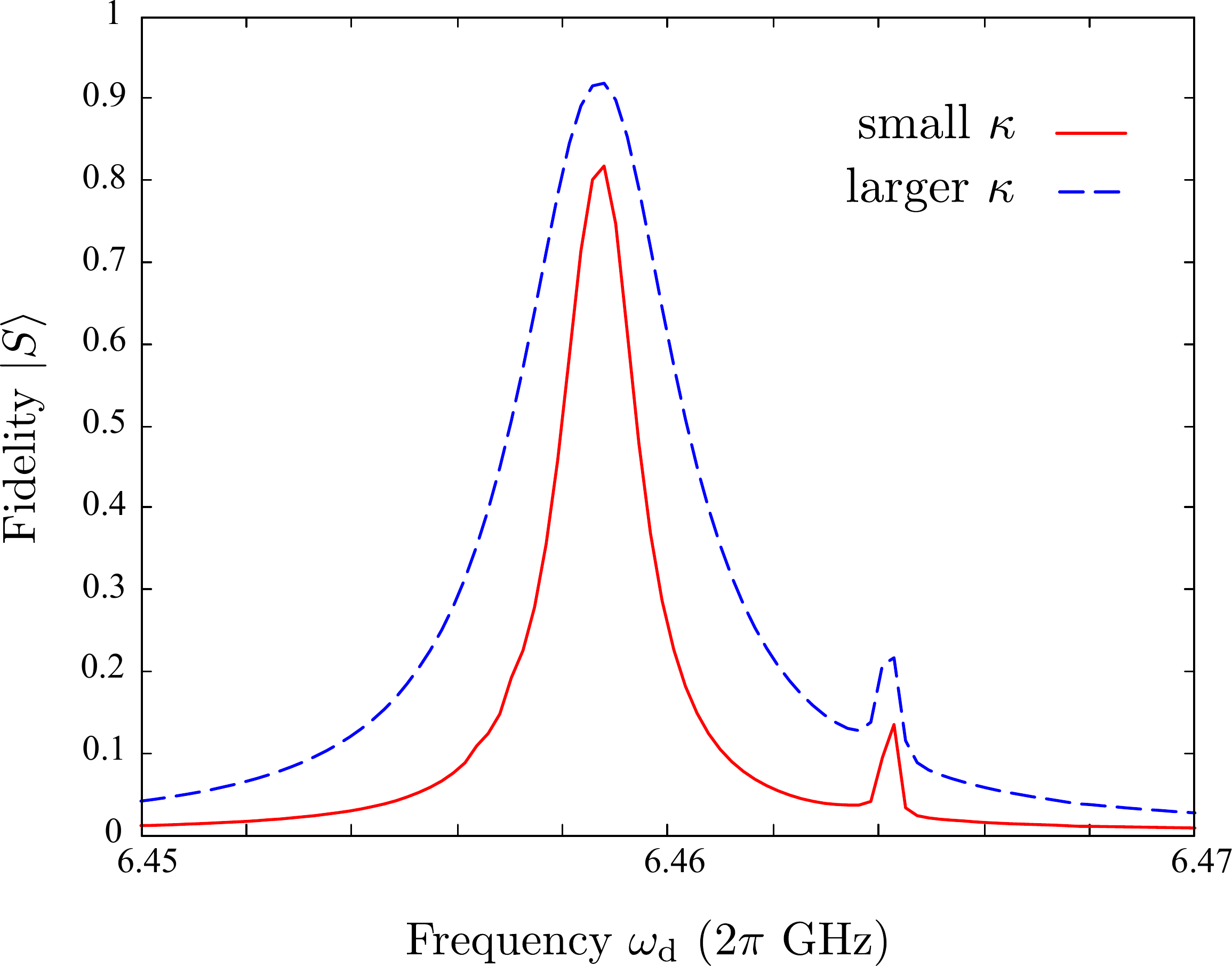}
 \caption{``Switching scheme'': steady-state fidelity to $\ket{S}$ as a function of $\omega_\rmd$. Calculations are performed using the approximate steady-state Liouvillian based on the time-independent Hamiltonian in Eq.~(\ref{eq:HamiltonianRot}). The results are shown for two different values of the cavity decay rate $\kappa$. Solid line: $\kappa=0.1\times 2\pi$ MHz; Dashed line: $\kappa=0.5\times 2\pi$ MHz. ($\gamma_\varphi = 5 \times 10^{-6}, \epsilon_\rmd = 0.25$, and $\epsilon_\rmd'=5\times 10^{-3}$ in units of $2\pi$~GHz).\label{fig:SwitchingResults}}
\end{figure}
We now use the steady-state approach of Sect.~IV to demonstrate the efficiency of the switching scheme. Targeting, for example, the singlet state $\ket{S}$, we use the first drive to take the qubit to $\ket{T_0}$ while the second drive is used to transfer the population to $\ket{S}$ with $\omega_\rmd'=\omega_\rmc^-+2J(g/\Delta)^2=6.102\times 2\pi$~GHz.
In Fig.~(\ref{fig:SwitchingResults}), we present the resulting steady-state fidelities to $\ket{S}$ as a function of the first drive frequency $\omega_\rmd$ for two different values of the cavity decay rate $\kappa$. A lossier cavity in this case leads to higher fidelities, reaching $92\%$. As we already discussed in Sect.~III, this can be explained by the fact that more lossy cavities favor a weaker intensity of anti-symmetric photonic fluctuations which in turn suppresses the backward process in which a cavity photon is removed.

\section{Scalability}

The presented purification scheme can be extended to larger systems with more than two qubits to realize generalized $W$-states with one excitation delocalized over the whole system~\cite{Aron2016}. Depending on the precise target state, it may require more than one additional drive. To illustrate this scalability property of the scheme, we discuss it for the case of three qubits with periodic boundary conditions. 
Given the geometry, both the photonic modes and the qubit eigenstates in the one-excitation manifold can be described by the quasi-momentum $k = 0, \pm 2\pi/3$. We choose as the target W-state for the qubits the fully symmetric $k=0$ spin-chain state $\ket{W}=(\ket{{\uparrow\downarrow\downarrow}}+\ket{{\downarrow\uparrow\downarrow}}+\ket{{\downarrow\downarrow\uparrow}})/\sqrt{3}$, while the remaining one-excitation eigenstates are $\ket{{\pm \frac{2\pi}{3}}}=(\ket{{\uparrow\downarrow\downarrow}}+\ehoch{\pm\frac{2\rmi\pi}{3}}\ket{{\downarrow\uparrow\downarrow}}+\ehoch{\mp\frac{2\rmi\pi}{3}}\ket{{\downarrow\downarrow\uparrow}})/\sqrt{3}$. These last two states are degenerate in energy, however they are separated from $\ket{W}$ by an energy difference $\delta$. Similarly, the cavity modes with $k=\pm 2\pi/3$ are degenerate with frequency $\omega_\rmc^{2\pi/3}$.

\begin{figure}
 \includegraphics[width=\linewidth]{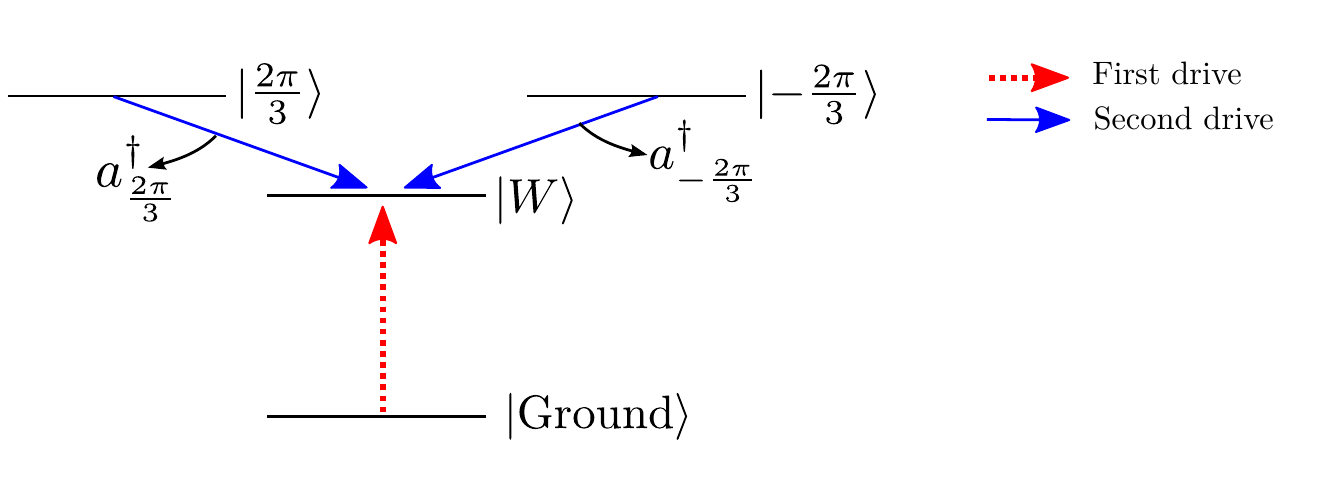}
 \caption{\label{fig:3level} (color online) How to stabilize a 3-qubit $W$ state: one drive (red, dotted arrow) takes the system from the ground state to the $W$ state, while the discussed scheme uses a second drive (blue, solid arrows) to fight dephasing to the states $\ket{{\pm\frac{2\pi}{3}}}$ via a resonant Raman process that scatters into the modes with wave vector $k=\pm\frac{2\pi}{3}$ described by $a^\dagger_{\pm\frac{2\pi}{3}}$. Note the degeneracy between $\ket{{\pm\frac{2\pi}{3}}}$ which allows us to use a single drive frequency to fight dephasing.
 }
\end{figure}

One possible implementation for our anti-dephasing scheme is the following (\textit{cf.} Fig.~\ref{fig:3level}): we pump each cavity in phase with a drive of frequency $\omega_\rmd'=\omega_\rmc^{2\pi/3}-\delta$. This triggers transitions from $\ket{{\pm\frac{2\pi}{3}}}$ to $\ket{W}$ by depositing a photon in the mode with $k=\pm\frac{2\pi}{3}$. Owing to the degeneracies, a single extra drive is required to counteract the dephasing mechanisms. For even larger systems, this may not be the case. To stabilize, \textit{e.g.}, a 5-qubit $W$ state in a ring of $5$ coupled cavities, symmetries require the use of two extra drives. We also see that some states cannot be addressed by the scheme. For example, due to the degeneracy, there is no possibility of stabilizing the $\ket{{\frac{2\pi}{3}}}$ state by driving from $\ket{{-\frac{2\pi}{3}}}$ to $\ket{{\frac{2\pi}{3}}}$.

\section{Conclusions}

We demonstrated a bath-engineering scheme to effectively counteract the dephasing mechanisms that limit the efficiency of the dissipative stabilization scheme discussed in Ref.~\cite{Aron2014}. The main ingredients are two driven-dissipative photonic modes with different parity and coupled to both qubits. The simplicity of this anti-dephasing scheme, together with its scalability to larger registers of qubits, makes it a promising approach in the ongoing efforts to achieve larger entangled states.

\begin{acknowledgments}
S.M.H. acknowledges support from Deutsche Forschungsgemeinschaft (DFG) through SFB 910 ``Control of self-organizing nonlinear systems'' (project B1) and through the ``School of Nanophotonics''. This work was supported by the U.S. Army Research Office (ARO) under grant no. W911NF-15-1-0299 and the NSF grant DMR-1151810.
\end{acknowledgments}
\appendix*

\section{Derivation of the transition rate presented in Eq.~(\ref{eq:gammatrans})\label{sc:rates}}
In this appendix, we give a detailed derivation of the transition rate created by the first ac drive.

We set $\epsilon_\rmd'=0$ to single out the first drive. First, we note that the term $\frac{g}{\Delta}\epsilon_\rmd\left(\sigma_1^x+\sigma_2^x\right)$ of $\widetilde{H}_{\text{S}}$ in Eqs.~(\ref{eq:hsys}) and~(\ref{eq:hx}) mixes the eigenstates of the fully un-driven $\widetilde{H}_{\text{S}}$ ($\epsilon_\rmd=\epsilon_\rmd'=0$)  presented in Eqs.~(\ref{eq:eigen}). The new states and their respective energy expectation values are, up to order $(g/\Delta)^2$,
\begin{align}
 \ket{\widetilde{T}_+}&\simeq\ket{T_+}+\frac{\Omega_\rmR}{\sqrt{2}\Delta_\rmq}\ket{T_0}& E_{\widetilde{T}_+}&\simeq\Delta_\rmq+\frac{\Omega_\rmR^2}{2\Delta_\rmq} \nonumber\\
 \ket{\widetilde{S}}&=\ket{S}&
 E_{\widetilde{S}}&=J(g/\Delta)^2\nonumber\\
 \ket{\widetilde{T}_0}&\simeq\ket{T_0}+\frac{\Omega_\rmR}{\sqrt{2}\Delta_\rmq}\big(\ket{T_-}-\ket{T_+}\big)&
 E_{\widetilde{S}}&=-J(g/\Delta)^2\nonumber\\
  \ket{\widetilde{T}_-}&\simeq\ket{T_-}-\frac{\Omega_\rmR}{\sqrt{2}\Delta_\rmq}\ket{T_0}& E_{\widetilde{T}_-}&\simeq-\Delta_\rmq-\frac{\Omega_\rmR^2}{2\Delta_\rmq}. \label{eq:newstates}
\end{align}
Here, we introduced $\Omega_\rmR\equiv2(g/\Delta)\epsilon_\rmd$ as well as ${\Delta_\rmq\equiv\omega_\rmq-\omega_\rmd +(g/\Delta)^2 \left[(\overline{N}_\rmd+1)\Delta+\sqrt{2}\mathrm{Re}(\epsilon_\rmd \overline{A}_\rmd)\right]}$. 

Between these states, transitions may be initiated through the coupling to the photon modes, which serve as a (structured) bath. These transitions are achieved through the terms in the Hamiltonian which couple to the photon noise operators $d$ or $D$, coming from
\begin{align}
  \widetilde{H}_\text{S-B}\equiv&\frac{1}{2}\left(\frac{g}{\Delta}\right)^2\Big[\Big(\overline{A}_\rmd \Delta+\frac{\epsilon_\rmd}{\sqrt{2}}\Big)D^\dagger\big(\sigma_1^z+\sigma_2^z\big)\nonumber\\
  &+\Big(\overline{A}_\rmd \Delta+\frac{\epsilon_\rmd}{\sqrt{2}}\Big)d^\dagger\big(\sigma_1^z-\sigma_2^z\big)\Big] +\text{H.c.} .
\end{align}
This part of the Hamiltonian can be written as
\begin{align}
 \widetilde{H}_\text{S-B}=\mathcal{S}^+ D^\dagger + \mathcal{S}^- d^\dagger +\text{H.c.},
\end{align}
where
\begin{align}
 \mathcal{S}^\pm\equiv&\frac{1}{2}\left(\frac{g}{\Delta}\right)^2\Big(\overline{A}_\rmd \Delta+\frac{\epsilon_\rmd}{\sqrt{2}}\Big)\big(\sigma_1^z\pm\sigma_2^z\big).
\end{align}

Invoking Fermi's Golden Rule, one can estimate the transition rates between the initial qubit state $\ket{i}$ and the final qubit state $\ket{f}$ by calculating
\begin{equation}
 \Gamma_{i\rightarrow f}=\Gamma^+_{i\rightarrow f}+\Gamma^-_{i\rightarrow f},
\end{equation}
where the transition matrix elements are
\begin{equation}
 \Gamma^\pm_{i\rightarrow f}\equiv2 \pi |\bra{f} \mathcal{S}^\pm \ket{i}|^2 \rho_\pm(E_i-E_f+\omega_\rmd). \label{eq:Fermi} 
\end{equation}

$\rho_\pm(\omega)$ is the density of states of the photonic bath in which a photon is emitted during the process, in our case provided by the two photonic modes, \textit{cf.} Eq.~(\ref{eq:DOS}). Since we are working in a frame rotating with $\omega_\rmd$, this frequency appears in the argument of $\rho_\pm$.   
Plugging in the states of Eq.~(\ref{eq:newstates}) into Eq.~(\ref{eq:Fermi}), one arrives at the transition rate given in Eq.~(\ref{eq:gammatrans}). Note that it is either the term coupling to $D$ or the term coupling to $d$ which is responsible to the transition, but never both. Therefore, for each transition, either $\Gamma^+_{i\rightarrow f}$ or $\Gamma^-_{i\rightarrow f}$ is identically zero. We find, as discussed above, that the transitions $\ket{\widetilde{T}_-}\rightarrow\ket{\widetilde{T_0}}$ and $\ket{\widetilde{T}_0}\rightarrow \ket{\widetilde{T}_+}$ involve the $D$ mode, leading to $\Gamma_{i\rightarrow f}=\Gamma^+_{i\rightarrow f}$. On the other hand, the transitions $\ket{\widetilde{T}_-}\rightarrow\ket{\widetilde{S}}$ and $\ket{\widetilde{S}}\rightarrow \ket{\widetilde{T}_+}$ involve the $d$ mode, leading to $\Gamma_{i\rightarrow f}=\Gamma^-_{i\rightarrow f}$. 

Concerning the Fermi's Golden rule, we want to point out that it is only applicable as long as the final state is not strongly populated. However, in the low-temperature limit discussed here, the ``final state'' always includes one excitation of a photonic mode, which quickly decays with the rate $\kappa$. Therefore, it is never strongly occupied, and Fermi's Golden Rule is applicable to calculate long-term steady state probability distributions. 

When the second drive is activated, $\epsilon_\rmd'\neq 0$, the values above actually change slightly. One needs to replace $\Delta_\rmq$ by
\begin{align}
 \Delta_\rmq'\equiv&\omega_\rmq-\omega_\rmd +(g/\Delta)^2 \Big[(\overline{N}_\rmd+\overline{N}_\rmd'+1)\Delta\nonumber\\
 &+\sqrt{2}\mathrm{Re}(\epsilon_\rmd \overline{A}_\rmd+\epsilon_\rmd' \overline{A}_\rmd')\Big].
\end{align}
This leads to minor changes only since the terms due to the second drive are much smaller than the terms due to the first drive when choosing $\epsilon_\rmd'\ll\epsilon_\rmd$.
\bibliography{antidephasingliterature_extended}

\end{document}